\documentclass[twocolumn,twoside,slac_two]{revtex4}
\usepackage{graphicx}
\usepackage{fancyhdr}
\usepackage{amsmath}
\usepackage{xspace}
\usepackage{listings}
\usepackage{multirow}
\usepackage[flushleft]{threeparttable}

\newcommand{\bbbar}[0]{\ensuremath{b \bar b}\xspace}
\newcommand{\sigmav}[0]{\ensuremath{\langle \sigma v \rangle}\xspace}
\newcommand{\TS}[0]{\ensuremath{\mathrm{TS}}\xspace}

\newcommand{\TSspec}[1]{\ensuremath{\mathrm{TS}_\mathrm{spec}^{#1}}\xspace}
\newcommand{\TSext}[1]{\ensuremath{\mathrm{TS}_\mathrm{ext}^{#1}}\xspace}

\newcommand{\unit}[1]{\ensuremath{\mathrm{\,#1}}\xspace}
\newcommand{\GeV}{\unit{GeV}}
\newcommand{\MeV}{\unit{MeV}}
\newcommand{\degree}{\ensuremath{^{\circ}}\xspace}
\newcommand{\cm}{\unit{cm}}

\newcommand{\second}{\unit{s}}
\newcommand{\photons}{\unit{ph}}

\newcommand{\code}[1]{\lstinline!#1!\xspace}
\newcommand{\Sourcelike}[0]{\lstinline!Sourcelike!\xspace}
\newcommand{\gtlike}[0]{\lstinline!gtlike!\xspace}

\newcommand{\gtobssim}[0]{\lstinline!gtobssim!\xspace}

\newcommand{\Figref}[1]{Figure~\ref{fig:#1}}
\newcommand{\Secref}[1]{Section~\ref{sec:#1}}

\newcommand{\Tabref}[1]{Table~\ref{tab:#1}}
\newcommand{\Eqnref}[1]{Eqn.~(\ref{eqn:#1})}

\begin{document}

\title{Search for Unknown Dark Matter Satellites of the Milky Way}

\author{Alex Drlica-Wagner\footnote{kadrlica@stanford.edu}, Ping Wang, Elliott Bloom, and Louis Strigari for the Fermi-LAT Collaboration}

\affiliation{W. W. Hansen Experimental Physics Laboratory, Kavli Institute for Particle Astrophysics and Cosmology, Department of Physics and SLAC National Accelerator Laboratory, Stanford University, Stanford, CA 94305, USA}

\begin{abstract}
We present a search for Galactic dark matter (DM) satellites using the Large Area Telescope (LAT). N-body simulations based on the $\Lambda$CDM model of cosmology predict a large number of as yet unobserved Galactic DM satellites.  These satellites could potentially produce $\gamma$-rays through the self-annihilation of DM particles.  Some DM satellites are expected to have hard $\gamma$-ray spectra, finite angular extents, and a lack of counterparts at other wavelengths. We searched for LAT sources with these characteristics.  We found no candidate DM satellites matching these criteria in one year of LAT data and interpreted this result in the context of N-body simulations.
\end{abstract}

\maketitle

\thispagestyle{fancy}

\section{Introduction} \label{sec:intro}

We report on a search for dark matter (DM) satellites via $\gamma$-ray emission from weakly interacting massive particle (WIMP) annihilation. Specifically, we searched for satellites predicted by cosmological N-body simulations~\citep{Diemand2007,Springel2008} but lacking counterparts in other wavelengths. We selected unassociated, high-Galactic-latitude $\gamma$-ray sources from both the First LAT Source Catalog (1FGL)~\citep{1FGL} and an independent list of source candidates created with looser assumptions on the source spectrum.  Using the likelihood ratio test, we distinguished extended sources from point sources and WIMP annihilation spectra from conventional power-law spectra. No candidates were found in either the unassociated 1FGL sources or our additional list of candidate sources.  This null detection is combined with the Via Lactea II (VL-II)~\citep{Diemand2007} and Aquarius~\citep{Springel2008} simulations to set an upper limit on the annihilation cross section for a $100\GeV$ WIMP annihilating through the \bbbar channel.

\section{Analysis} \label{sec:methods}

\subsection{Data Selection}
\label{subsec:data}
Our data sample consisted of `Diffuse' class events with energies between $200\MeV$ and $300\GeV$ from the first year of LAT data collection (2008 August 8 -- 2009 August 7). We rejected events with zenith angles larger than 105\degree and events taken during time periods when the rocking angle of the LAT was greater than 47\degree.  This analysis was limited to sources with Galactic latitudes greater than 20\degree, since the Galactic diffuse emission complicates source detection and the analysis of spatial extension at lower Galactic latitudes.  We modeled the diffuse $\gamma$-ray emission with the standard Galactic (\textit{gll\_iem\_v02.fit}) and isotropic (\textit{isotropic\_iem\_v02.txt}) background models.  Throughout this analysis, we used the LAT ScienceTools version v9r18p1 and the P6\_V3\_DIFFUSE instrument response functions.\footnote{http://fermi.gsfc.nasa.gov/ssc/data/}

\subsection{Source Selection}
\label{sec:search}
The 1FGL contains 1451 high-energy $\gamma$-ray sources, of which 806 are at high Galactic latitude ($|b|>20\degree$).  Of these high-latitude 1FGL sources, 231 are unassociated with sources at other wavelengths and constitute the majority of the sources that were examined as potential DM satellites.  The 1FGL spectral analysis, including the threshold for source acceptance, assumed that sources were point-like with power-law spectra. This decreased the sensitivity of the 1FGL to both spatially extended and non-power-law sources, which are characteristics expected for DM satellites. To mitigate these biases, we augmented the unassociated sources in the 1FGL with an independent search of the high-latitude sky~\citep{Wang:2011}.

We searched for $\gamma$-ray sources using the internal LAT Collaboration software package, \Sourcelike~\citep{Abdo:2010qd}.  \Sourcelike performs a fully binned likelihood fit in two dimensions of space and one dimension of energy. For spectral fitting, \Sourcelike fits the number of counts associated with a source in each energy bin independently. The full likelihood is the product of the likelihoods in each bin.  This calculation has more degrees of freedom than that performed by the LAT ScienceTool, \gtlike, which calculates the likelihood from all energy bins simultaneously according to a user-supplied spectral model. In this analysis, we used 11 energy bins logarithmically spaced from $200\MeV$ to $300\GeV$.

Using \Sourcelike, we searched for sources in 2496 regions of interest (ROIs) of dimension $10\degree \times 10\degree$ centered on HEALPix~\citep{Gorski:2005} pixels obtained from an order 4 tessellation of the high-latitude sky ($|b| > 20\degree$).  Each  ROI was sub-divided into $0.1\degree \times 0.1\degree$ pixels, and for each pixel the likelihood of a point source at that location was evaluated by comparing the maximum likelihood ($\mathcal{L}$) of two hypotheses: (1) that the data were described by the standard LAT diffuse background models without any point sources ($H_0$), and (2) that the data were described by the existing model with an additional free parameter corresponding to the flux of a source at the target location ($H_1$).  Utilizing the likelihood ratio test, we defined a test statistic, $\TS = -2 \ln ({\mathcal{L}(H_0)}/{\mathcal{L}(H_1)})$.

After creating a map of \TS over the high-latitude sky, candidate sources with $\TS > 16$ were iteratively refit incorporating the normalizations of diffuse backgrounds and neighboring point sources as free parameters.  After refitting, source candidates with $\TS > 24$\footnote{Simulations show that 1 in $10^4$ background fluctuations are detected at $\TS \geq 24$ when fit with \Sourcelike.~\citep{Wang:2011}.} were accepted for study. Finally, to avoid duplicating 1FGL sources, we removed candidate sources with 68\% localization errors overlapping the 95\% error ellipse given in the 1FGL.

Our search of the high-latitude sky revealed 710 candidate sources, of which 154 are not in the 1FGL (36 of these candidate sources were subsequently included in the Second LAT Source Catalog (2FGL)~\citep{2FGL}).  We did not expect to recover all of the 806 high-latitude 1FGL sources, since the 1FGL is a union of four different detection methods and external seeds from the BZCAT and WMAP catalogs~\citep{1FGL}.  However, since \Sourcelike fits each energy bin independently, we expected to find source candidates that were not included in the 1FGL, either because they had non-power-law spectra or they had hard spectra with too few photons to pass the 1FGL spectral analysis. We sacrificed some sample purity for detection efficiency in our candidate source list because stringent cuts on spatial extent and spectral shape were later applied.  We obtained a final list of 385 high-latitude unassociated LAT sources and source candidates by combining 231 unassociated 1FGL sources with 154 non-1FGL candidate sources.

\begin{figure}[t]
\centering
\includegraphics[width=\columnwidth]{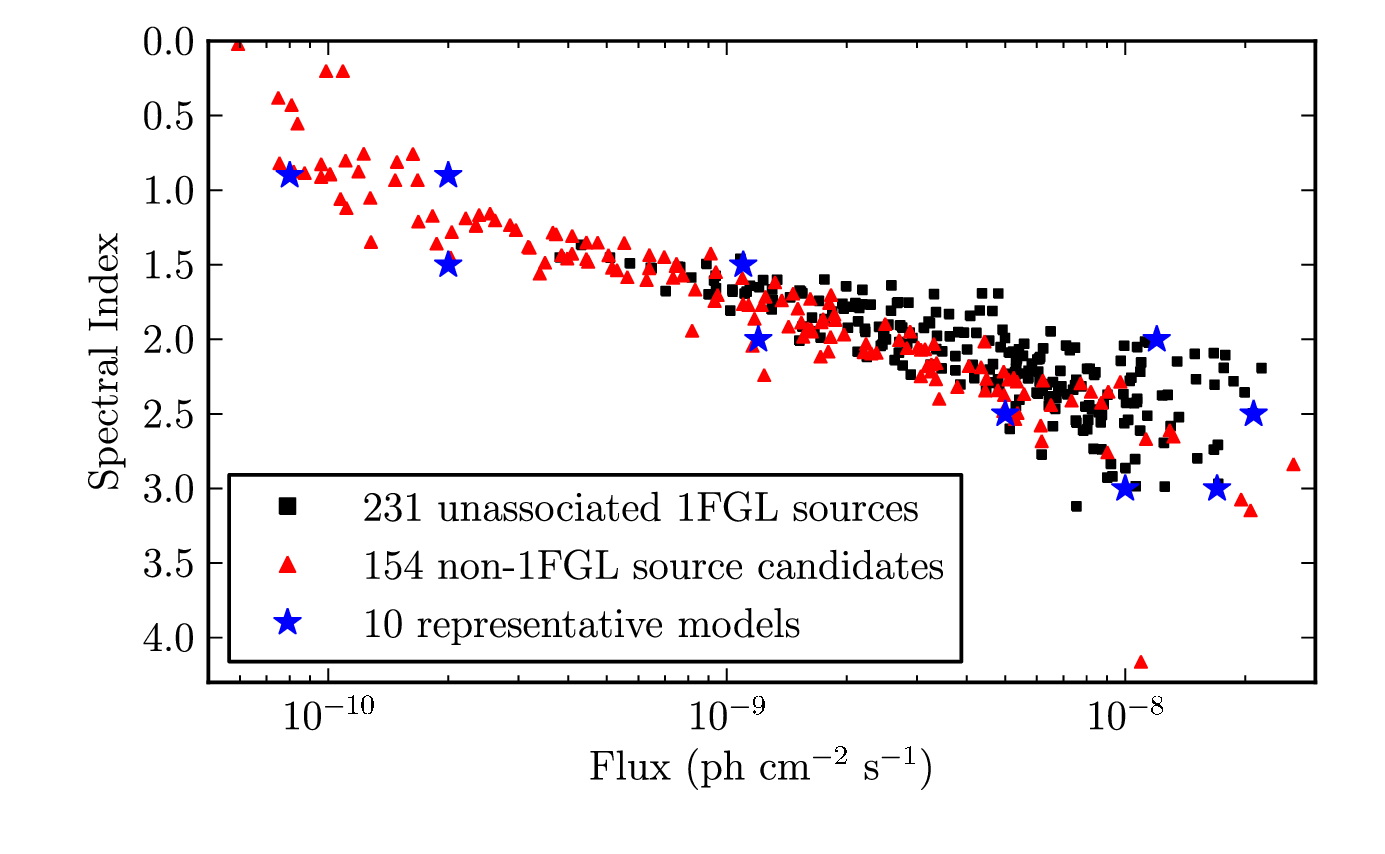}
\caption{The distribution of spectral indices and integral fluxes from $200\MeV$ to $300\GeV$ for the 385 high-latitude unassociated sources and source candidates. The squares are the 231 unassociated sources from the 1FGL, while the triangles are the 154 additional source candidates detected with \Sourcelike.  The stars are the 10 representative power-law models in \Tabref{tests}.} \label{fig:image1}
\end{figure}

To check for consistency with the source analysis of the 1FGL, we performed an unbinned likelihood analysis with \gtlike assuming that the unassociated sources were point-like with power-law spectra.  Our fitted fluxes and spectral indices are in good agreement with those in the 1FGL for the 231 unassociated 1FGL sources. The wide range of fluxes and spectral indices spanned by the unassociated LAT sources can be seen in \Figref{image1}. It is apparent that there are more non-1FGL source candidates in this sample with very hard spectra (spectral index $ \sim 1.0$) and very low fluxes ($\sim 10^{-10} \photons \cm^{-2} \second^{-1}$). We find that these source candidates are very likely spurious~\citep{Wang:2011}.

\subsection{DM Satellite Candidate Selection}
\label{subsec:exttest}
The $\gamma$-ray flux from WIMP annihilation in a satellite can be expressed as
\begin{align}
\label{eqn:flux}
\phi_{\mathrm{WIMP}}(E,\psi) = J(\psi) \times \Phi^{\mathrm{PP}}(E).
\end{align}
Here, $J(\psi)$ represents the line-of-sight integral through the DM density at an offset angle $\psi$ (relative to the center of the satellite), while $\Phi^{\mathrm{PP}}(E)$ is an energy-dependent particle physics factor including the velocity-averaged annihilation cross section \sigmav (see~\citep{Bergstrom:1997fg} for more details).

While the majority of DM satellites in current simulations are not spatially resolvable by the LAT, spatial extension is an important feature for distinguishing large or nearby satellites from point-like astrophysical sources.  Assuming that the DM distribution of a satellite follows an NFW profile~\citep{Navarro:1996gj} with scale radius $r_s$ at a distance $D$, the angular extent of the satellite can be characterized by the parameter $\alpha_0 = r_s/D$. Approximately $90\%$ of the integrated $J$-factor comes from within the angular radius $\alpha_0$~\citep{Strigari:2006rd}.

We used the likelihood ratio test, as implemented by \Sourcelike, to test sources for spatial extension. We defined a test statistic for extension as $\TSext{} =  \TS_\mathrm{NFW}-\TS_\mathrm{point}$, where $\TS_\mathrm{point}$ is the test statistic of the candidate source assuming that it has negligible extension ($\alpha_0$ much smaller than the LAT PSF) and $\TS_\mathrm{NFW}$ is the test statistic of the candidate source when $\alpha_0$ is fit as a free parameter.  In both cases, the position of the source was optimized during the fit.

Additionally, we defined a test statistic to select sources with $\gamma$-ray spectra consistent with WIMP annihilation into \bbbar (chosen as a representative proxy for tree-level annihilation spectra).  This spectral test statistic, $\TSspec{} = \TS_{\bbbar} - \TS_\mathrm{pwl}$, is the difference in source \TS calculated with an unbinned analysis using \gtlike assuming a \bbbar spectral model ($\TS_{\bbbar}$) and a power-law ($\TS_\mathrm{pwl}$) spectral model. When performing our fit, we modeled the candidate source as point-like\footnote{Assuming that sources are point-like was found to be a conservative way to estimate $\TSspec{}$~\citep{Wang:2011}.} and left its flux and the flux of the diffuse backgrounds free. In addition to the flux, each spectral model contained an additional free parameter (the DM mass or spectral index).

We defined cuts to independently eliminate 99\% of point sources and 99\% of power-law sources, labelled \TSext{99} and \TSext{99} respectively. To evaluate these cuts over the pertinent range of source fluxes and spectral indices, we bracketed the unassociated LAT sources with 10 representative power-law models (\Tabref{tests} and blue stars in \Figref{image1}).  For each of these 10 models, we simulated 1000 independent sources at random locations in the high-latitude sky using the LAT simulation tool, \gtobssim, and the spacecraft pointing history for our one-year data set.  To accurately incorporate imperfect modeling of background point and diffuse sources, we embedded the simulated point sources in the LAT data and calculated \TSext{} and \TSspec{}.  We defined \TSext{99} and \TSspec{99} for each representative model as the smallest value of \TSext{} or \TSspec{} that was larger than that calculated for 99\% of simulated power-law point sources.  The values of \TSext{99} and \TSspec{99} (\Tabref{tests}) were calculated independently of each other and the $\TS > 24$ detection cut. We used a bilinear interpolation of \Tabref{tests} to estimate the value of \TSext{99} and \TSspec{99} for any point in the space spanned by the grid of flux and spectral index. These tests of spatial extension and spectral character allow us to select non-point-like and non-power-law sources with a contamination of 1 in $10^{4}$ assuming they are independent.
\begin{table}[t]
\caption{Values for \TSext{99} and \TSspec{99}}
\begin{threeparttable}
\begin{tabular}{ccccc}
\hline
\hline
Model & Spectral & Flux\tnote{(a)}             & \TSext{99} & \TSspec{99}  \\
Number& Index    & $(\photons \cm^{-2} \second^{-1})$ &                      & \\
\hline
1 & 0.9 & $2.0\times {10}^{-10}$ &  6.18  & 2.38 \\  
2 & 0.9 & $8.0\times {10}^{-11}$ &  7.87  & 2.46 \\  
3 & 1.5 & $1.1\times {10}^{-9}$  &  5.09  & 4.96 \\
4 & 1.5 & $2.0\times {10}^{-10}$ &  14.98 & 2.88 \\ 
5 & 2.0 & $1.2\times {10}^{-8}$  &  5.11  & 2.24 \\ 
6 & 2.0 & $1.2\times {10}^{-9}$  &  9.63  & 4.28 \\ 
7 & 2.5 & $2.1\times {10}^{-8}$  &  6.74  & 1.78 \\ 
8 & 2.5 & $0.5\times {10}^{-8}$  &  10.78 & 5.66 \\  
9 & 3.0 & $1.7\times {10}^{-8}$  &  9.81  & 2.14 \\ 
10& 3.0 & $1.0\times {10}^{-8}$  &  11.87 & 6.02 \\  
\hline
\end{tabular}
\begin{tablenotes}
   \item ${}^{\rm (a)}$Integral flux from $200\MeV$ to $300\GeV$.
   \item Note -- Cuts for excluding point-like sources and sources with power-law spectra. 
\end{tablenotes}
\end{threeparttable}
\label{tab:tests}
\end{table}

\section{Results} \label{results}
We applied the cuts on spatial extension and spectral character to select DM satellite candidates from the 385 unassociated high-latitude LAT sources and source candidates. Two of the 385 unassociated sources, 1FGL\,J1302.3$-$3255 and 1FGL\,J2325.8$-$4043, passed the cut on spatial extension.  One of these, 1FGL\,J1302.3$-$3255, also passed our spectral test, preferring a \bbbar spectrum to a power-law spectrum.  However, we do not believe that either of these sources is a viable DM satellite candidate for reasons discussed below.

1FGL\,J1302.3$-$3255 was unassociated when the 1FGL was published but has since been associated with a millisecond pulsar by radio follow-up observation~\citep{Hessels:2011vc}. The other source, 1FGL\,J2325.8$-$4043, has a high probability of association with two AGN~\citep{Abdo:2010ge}, though it did not meet the association criteria of the 1FGL. Cross checking against the 2FGL, two sources were found within 0.5\degree of the location of 1FGL\,J2325.8$-$4043~\citep{2FGL}. In one year of data, these two sources could not be spatially resolved, but their existence was enough to favor an extended source hypothesis.

Since 1FGL\,J1302.3$-$3255 was associated with a pulsar and 1FGL\,J2325.8$-$4043 did not appear to be truly extended, we conclude that there were no unassociated, high-latitude spatially extended $\gamma$-ray sources in the first year of LAT data. Thus, according to the criteria defined in \Secref{methods}, no viable DM satellite candidates were found.

\section{Discussion}
Using the detection efficiency of our selection, the absence of DM satellite candidates can be combined with the Aquarius and VL-II simulations to constrain a conventional $100\GeV$ WIMP annihilating through the \bbbar channel. We calculated the probability of detecting no satellites from the individual detection efficiency of each satellite in the realization. By increasing the satellite flux until the probability of detecting no satellites drops below 5\%, we set a $95\%$ confidence upper limit on \sigmav.

\subsection{Detection Efficiency}
\label{subsec:efficiency}
The detection efficiency of our selection is defined as the fraction of true DM satellites that would pass the cuts in \Secref{methods} and was calculated from Monte Carlo simulations.  The efficiency for detecting a DM satellite depends on spectral shape (i.e., DM mass and annihilation channel), flux, and spatial extension.  For a $100\GeV$ WIMP annihilating through the \bbbar channel, we examined the efficiency for satellites with fluxes in the range of the unassociated high-latitude LAT sources and angular extents in the range to which the LAT is currently sensitive (\Tabref{bbbar}).

For each set of characteristics listed in \Tabref{bbbar}, we simulated 200 DM sources with NFW profiles and \bbbar spectra from a $100\GeV$ WIMP.  These simulations were embedded in LAT data at random high-latitude locations, and \Sourcelike was used to compute \TSext{}, \TSspec{}, and the detection \TS for each.  The satellite detection efficiency was computed as the fraction of satellites with \Sourcelike $\TS > 24$, $\TSext{} > \TSext{99}$, and $\TSspec{} > \TSspec{99}$.  The first requirement was included as a proxy for the efficiency of the source finding algorithm. To expedite the generation of this table, we found the flux value with efficiency $<0.05$ and conservatively set the efficiency for sources with less flux to 0.
\begin{table}[t]
\caption{Detection efficiency}
\begin{threeparttable}
\begin{tabular}{ccccc}
\hline
\hline
Flux\tnote{(a)}                    &   &            & Extension &  \\
($\photons \cm^{-2} \second^{-1}$)  &   & 0.5\degree & 1.0\degree & 2.0\degree \\
\hline
$0.2\times 10^{-8}$   &   & $<0.05$   &   $<0.05$   & $<0.05$     \\
$0.5\times 10^{-8}$   &   & 0.16      &   0.28      & 0.31        \\    
$1.0\times 10^{-8}$   &   & 0.74      &   0.76      & 0.83        \\
$2.0\times 10^{-8}$   &   & 0.99      &   1.0       & 0.99        \\  
$5.0\times 10^{-8}$   &   & 1.0       &   1.0       & 1.0         \\
\hline
\end{tabular}
\begin{tablenotes}
   \item ${}^{\rm (a)}$Integral flux from $200\MeV$ to $300\GeV$.
   \item Note -- Satellite detection efficiency for a $100\GeV$ WIMP annihilating to \bbbar.
\end{tablenotes}
\end{threeparttable}
\label{tab:bbbar}
\end{table}

\subsection{Upper Limits}
\label{subsec:upperlimits}

Realizations of the Galactic DM satellite population were created by selecting 6 maximally separated vantage points 8.5 kpc from the center of the VL-II simulation and each of the 6 Aquarius simulations. For each of the $6 \times (1 + 6) = 42$ ``visualizations'' of  VL-II and Aquarius, we calculated the $\gamma$-ray fluxes of all satellites for a given \sigmav using \Eqnref{flux}. With these fluxes and the true spatial extension for each satellite, we performed a bilinear interpolation on \Tabref{bbbar} to determine the detection efficiency for each satellite. The probability that the LAT would observe none of the satellites in visualization $i$ is
\begin{equation}
\label{eqn:prob}
P_i( \sigmav ) = \prod_{j} (1 - \epsilon_{i,j}( \sigmav ))
\end{equation}
where $\epsilon_{i,j}$ is the detection efficiency for satellite $j$ in visualization $i$. Because there is no reason to favor any one visualization, we calculated the average null detection probability over the $N=42$ visualizations as
\begin{equation}
\bar P (\sigmav) = \frac{1}{N} \sum_i^N P_i( \sigmav )
\end{equation}
To set an upper limit on the DM annihilation cross section, we increased \sigmav until the probability of a null observation was $<5\%$, i.e. $\bar P < 0.05$. This corresponds to 95\% probability that, for this \sigmav, at least one satellite would have passed our selection criteria. Using this methodology, the LAT null detection constrains \sigmav to be less than $1.95 \times 10^{-24} \cm^{3} \second^{-1}$ for a $100\GeV$ WIMP annihilating through the \bbbar channel.

\begin{acknowledgments}
The $Fermi$ LAT Collaboration acknowledges support from a number of agencies and institutes for both development and the operation of the LAT as well as scientific data analysis. These include NASA and DOE in the United States, CEA/Irfu and IN2P3/CNRS in France, ASI and INFN in Italy, MEXT, KEK, and JAXA in Japan, and the K.~A.~Wallenberg Foundation, the Swedish Research Council and the National Space Board in Sweden. Additional support from INAF in Italy and CNES in France for science analysis during the operations phase is also gratefully acknowledged.

ADW is supported in part by the Department of Energy Office of Science Graduate Fellowship Program (DOE SCGF), made possible in part by  the American Recovery and Reinvestment Act of 2009, administered by ORISE-ORAU under contract no. DE-AC05-06OR23100.
\end{acknowledgments}


\end{document}